\documentclass[a4paper,11pt]{article}
\pdfoutput=1 

\usepackage{jinstpub} 
\usepackage{hyperref}

\usepackage{todonotes}  
\usepackage{subfig} 

\graphicspath{{./figures/}}

\title{Electrical breakdown in Thick-GEM based WELL detectors}

\author[a,1]{A. Jash \note{Corresponding author.}}
 \author[a]{L. Moleri,}
 \author[a]{S. Bressler}

\affiliation[a]{Department of Particle Physics and Astrophysics, Weizmann Institute of Science, Herzl st. 234, Rehovot 7610001, Israel}

\emailAdd{abhik.jash@weizmann.ac.il}
\abstract
{
The occurrence of electrical discharges in gas detectors restricts their dynamic range and degrades their performance. Among the different methods developed to mitigate discharge effects, the use of resistive materials in the detector assembly was found to be very effective. In this work, we present the results of a comparative study of electrical discharges in Thick-GEM-based WELL-type detectors - with and without resistive elements. We present a new method to measure discharges in the resistive-detector configurations; it allows demonstrating, for the first time, the occurrence of discharges also in the Resistive-Plate WELL detector configuration. It also provides direct evidence for the Raether limit.
}
\begin{document}
\maketitle
\flushbottom
\section{Introduction}
\label{section:introduction}
%
Micro-Pattern Gas Detectors (MPGDs) are candidates of choice for many ongoing and future Particle-physics experiments due to their good spatial and energy resolutions, high detection efficiency, high rate capabilities, and the potential to be industrially produced. Nonetheless, the dynamic range of MPGDs is often limited by the occurrence of occasional discharges that could result in efficiency loss, dead time, and damage to the detector electrodes and the readout electronics. 

The sequence of events leading to a discharge is initiated when the avalanche size exceeds a critical charge limit ($10^{6} - 10^{7}$ ion-electron pairs) - the so-called Raether limit \cite{Raether}. The resulting local electric field becomes large enough to induce a transition of the avalanche to a forward-backward propagating streamer, a well-studied process (see for example \cite{Raether, fonte_peskov, fonte_streamers, peskov_feedback_breakdown, Battistoni_wireChamber, Taylor_wireChamber}). Due to the small distance between the electrodes in most MPGD configurations, the streamer is likely to form an electrical connection between neighboring electrodes of different potentials, consequently discharging the energy stored in the equivalent capacitor.

Different methods were developed over the years to resolve the problem of discharges. E.g., in cascaded-electrodes structures \cite{amos_multistepChamber, Bressan_discharge_limit, Charles_discharge_MM} the charge density at each stage is reduced, and higher gains could be reached prior to the onset of discharges. Dividing the detector into smaller segments does not allow higher gains; instead, it reduces the area affected by a discharge and the corresponding discharge energy stored in the involved capacitor \cite{Bachmann_discharge_GEM}. In recent years, the most common approach for mitigating discharges in MPGDs has been embedding resistive electrodes in the detector assembly \cite{dixit_resistiveAnode, Alexopoulos_resistiveMM, Mauro_RETGEM, Rui_kaptonTHGEM, abbrescia_book, yoshikawa_REGEM, resistiveMPGD_advances, Bencivenni_uRWELL}. This has two roles: (i) Protecting the readout electronics by decoupling it from the energy released in the discharge. (ii) Quenching the discharge energy. As shown for Resistive Plate Chambers \cite{RPC}, the long clearance time of charges from the resistive-electrode results in a local reduction of the electric field and self-extinction of the discharge.

In the present work, we study discharges in three different Thick Gas Electron Multiplier \cite{THGEM1, THGEM2} (THGEM)-based WELL detectors: standard THick-WELL (THWELL) \cite{Lior_DHCAL} with no resistive material embedded, Resistive-WELL (RWELL) \cite{Lior_DHCAL}, and Resistive Plate WELL (RPWELL) \cite{RPWELL}. The three configurations are discussed below. We demonstrate in this work the effect of the resistive materials on various discharge characteristics.

The THGEM electrode is a scaled-up ($\sim$tenfold) variant of the Gas Electron Multiplier (GEM) \cite{GEM} foil. They are produced by the standard printed circuit board (PCB) technology: mechanical drilling of sub-millimeter diameter holes pattern in insulating (e.g., FR4) plates, Cu-clad on both sides, followed by chemical etching of concentric insulating rims around the hole edges. The latter was found to considerably reduce the discharge probability~\cite{Shikma_instabilities_THGEM}, at the cost of charging up effects \cite{chargingUp_michael, gainStabilization_dan}. Typical THGEM electrodes have $\sim$0.5~mm thickness, 0.5~mm in diameter holes, $\sim$1~mm hole pitch, and $\sim$0.1~mm hole rim.

THGEM-based detectors are suitable for applications requiring large-area coverage at sub-mm spatial resolution \cite{Cortesi_THGEM_submm, Luca_positionResolution_RPWELL}, a few ns time resolution \cite{THGEM_timeRes}, high counting rate capability \cite{Shalem_THGEM_advances} and robustness against occasional discharges \cite{Chechik_advances_GPM}.

In a standard THGEM-detector configuration, the THGEM electrode is preceded by a conversion drift gap. Radiation-induced ionization electrons drift into the THGEM holes, where multiplication occurs under a high electric field. The resulting avalanche electrons are extracted into a few-mm wide induction gap and drift towards the readout anode, inducing an electrical signal.

The WELL electrode is a THGEM electrode Cu-clad on one side only. In a THWELL detector configuration (Figure \ref{fig:schema_WELL}), the WELL electrode is coupled directly to the anode (typically, a Cu-clad FR4). The absence of the induction gap leads to a significantly thinner geometry - an advantage for applications with stringent space limitations \cite{shikma_RPWELL_SDHCAL}. Compared to a THGEM with an induction gap, higher gains could be obtained for lower applied voltage across the WELL electrode due to the larger electric field within the closed holes \cite{Lior_WELLs}. Signals are induced on the readout anode by the movement of avalanche electrons and ions within the holes. 

In the RWELL detector \cite{Lior_DHCAL} (Figure \ref{fig:schema_RWELL}), the WELL-electrode is coupled to the readout anode through a resistive layer, deposited on an insulating sheet. Charges reaching the resistive layer propagate sideways and evacuate to the ground. This configuration was proposed in an attempt to protect the multiplier, its anode, and the readout electronics from discharge damages and to minimize dead-time effects following a discharge. Using resistive material of 10 M$\Omega/\square$ surface resistivity, it was shown that while the discharge probability remains similar to that of a THWELL configuration \cite{Lior_DHCAL, Shikma_instabilities_THGEM}, the energy released during a discharge is, indeed, quenched by an order of magnitude. 

In the RPWELL \cite{RPWELL} detector (Figure \ref{fig:schema_RPWELL}), the WELL electrode is coupled to the readout anode through a material of high bulk resistivity ($\sim$10\textsuperscript{9} - 10\textsuperscript{12} $\Omega$cm). 
Charges reaching the resistive plate are evacuated to the ground through the bulk of the resistive material. The typical resistance to ground in an RPWELL configuration is higher than in an RWELL, which should provide superior discharge protection and quenching. Indeed, in earlier works \cite{RPWELL, Luca_RPWELL_Ar, Shikma_RPWELL_in-beam, Shikma_THGEM_advances}, in which discharges were defined as rather large ($\mathrm{>10~nA}$) current fluctuations, the RPWELL has always been treated as a discharge-free detector. 

In the present work, we demonstrate the appearance of discharges in RPWELL detectors as well and characterize them relative to discharges occurring in the THWELL and RWELL configurations. In section \ref{section:expt_setup}, we describe the investigated detector configurations, the experimental setup and the methodology; the results comparing the three configurations in terms of discharge intensity and probability are presented in section \ref{section:results}, followed by a discussion in section \ref{section:summary}.
\section{Experimental setup and Methodology}
\label{section:expt_setup}
\subsection{WELL structures}
\label{section:well_structures}
\begin{figure}[htb]
 \centering
   \subfloat[a][]{
   \centering
    \includegraphics[width=0.3\textwidth]{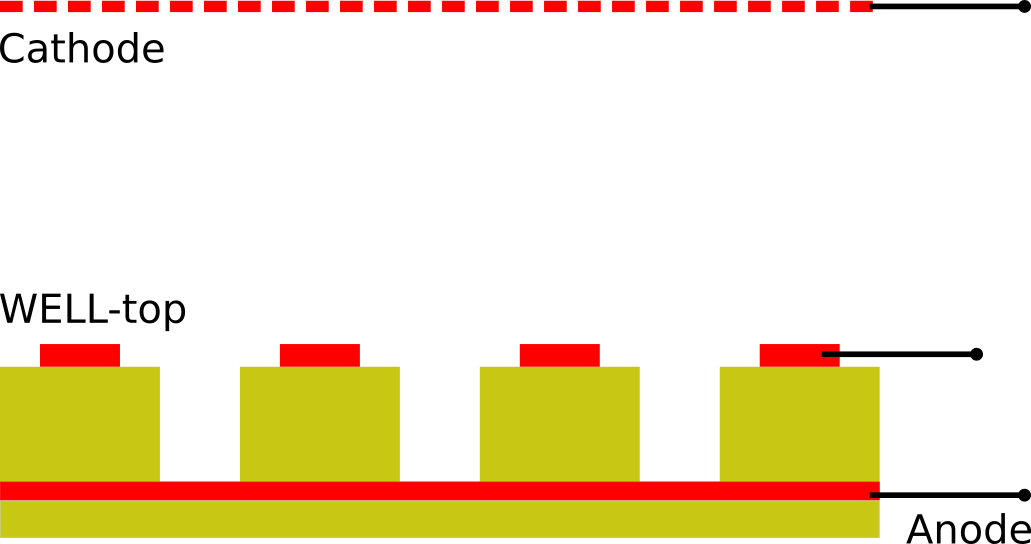}
    \label{fig:schema_WELL}
    }
 \hfill
   \subfloat[b][]{
   \centering
    \includegraphics[width=0.3\textwidth]{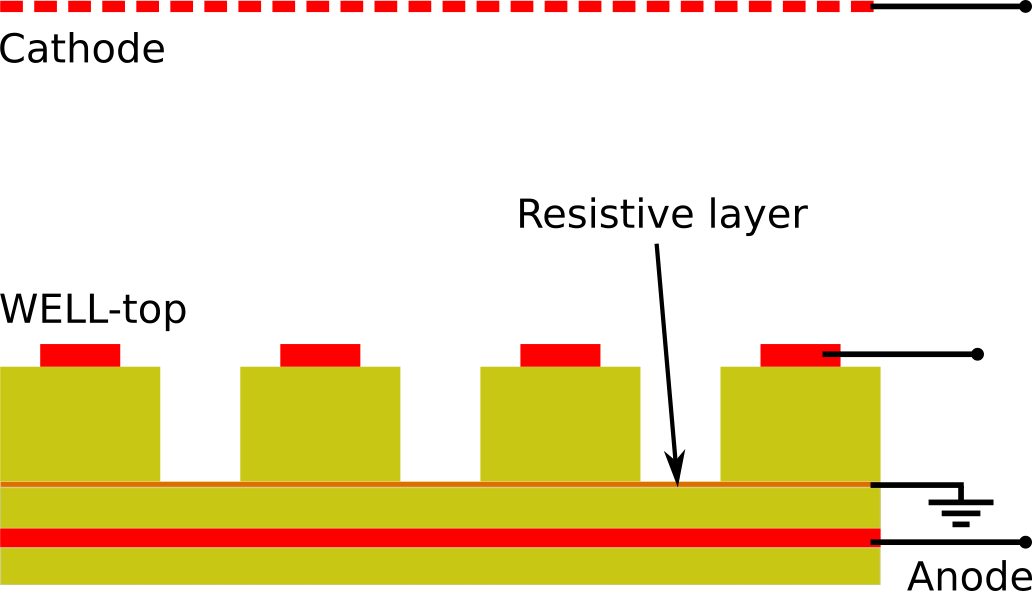}
    \label{fig:schema_RWELL}
    }
 \hfill
   \subfloat[c][]{
   \centering
    \includegraphics[width=0.3\textwidth]{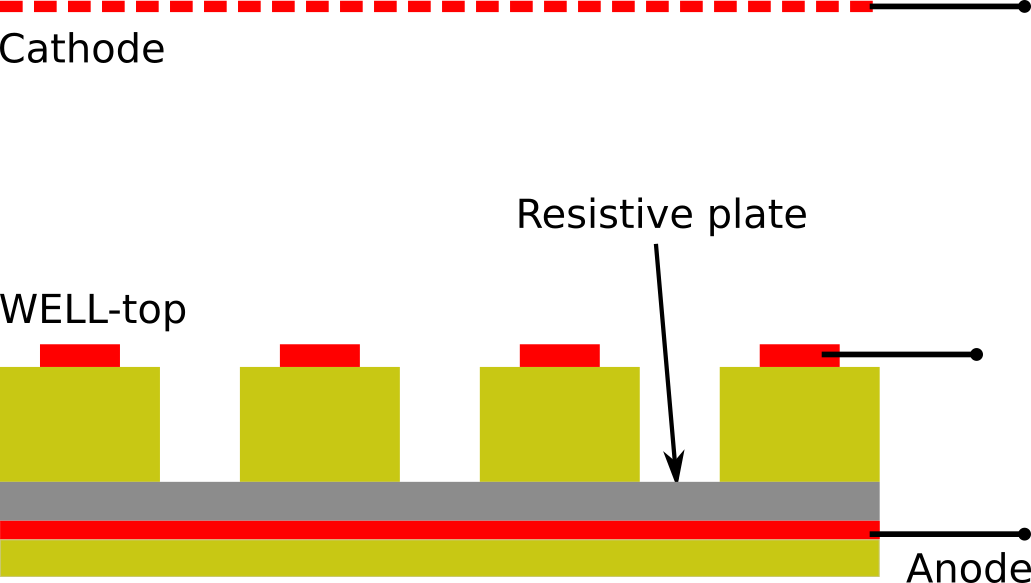}
    \label{fig:schema_RPWELL}
    }
 \caption{Schematic diagram (not to scale) and bias scheme of the three detector types: \protect\subref{fig:schema_WELL} THWELL, \protect\subref{fig:schema_RWELL} RWELL, \protect\subref{fig:schema_RPWELL} RPWELL.}
 \label{fig:schema_3Det}
\end{figure}

For all the studied configurations, we used a 0.4~mm thick single-faced THGEM electrode of 20$\times$20 mm\textsuperscript{2} active area. It has 0.5 mm diameter holes with 0.1 mm rim, arranged in a square pattern with 1 mm hole pitch. A Stainless steel mesh was used as the cathode, located 3 mm above the WELL electrode. The mesh allows the passage of low-energy X-rays for the detector characterization. The three WELL-based detector configurations are described below.
\begin{enumerate}
 \item \textbf{THWELL} - the WELL electrode is directly coupled to a Cu-clad FR4 anode (Figure \ref{fig:schema_WELL}).
 \item \textbf{RWELL} - a mixture of graphite and lacquer providing a surface resistivity of 16 M$\Omega/\square$ was sprayed on an FR4 plate. Two such plates of 0.9~mm and 0.2~mm thickness were used.
 This resistive electrode was placed between the WELL electrode and the anode - with its graphite-painted side facing the THGEM (Figure \ref{fig:schema_RWELL}). To allow charge evacuation through the graphite layer, one of its edges was connected to the ground by a Cu line. 
 \item \textbf{RPWELL} - a 0.7 mm thick low-resistive silicate (LRS) glass \cite{LRS_glass} with bulk resistivity 2 $\times$ 10\textsuperscript{10} $\Omega$cm was placed in between the WELL electrode and anode (Figure \ref{fig:schema_RPWELL}). The bottom face of the resistive glass is attached to the anode using a conductive tape \cite{3M_tape} to ensure good electrical contact between the two.
 \end{enumerate}
\subsection{Setup}
\label{section:setup}
\begin{figure}[!htbp]
\centering
	\includegraphics[width=0.95\textwidth, trim={0 0 0 0},clip]{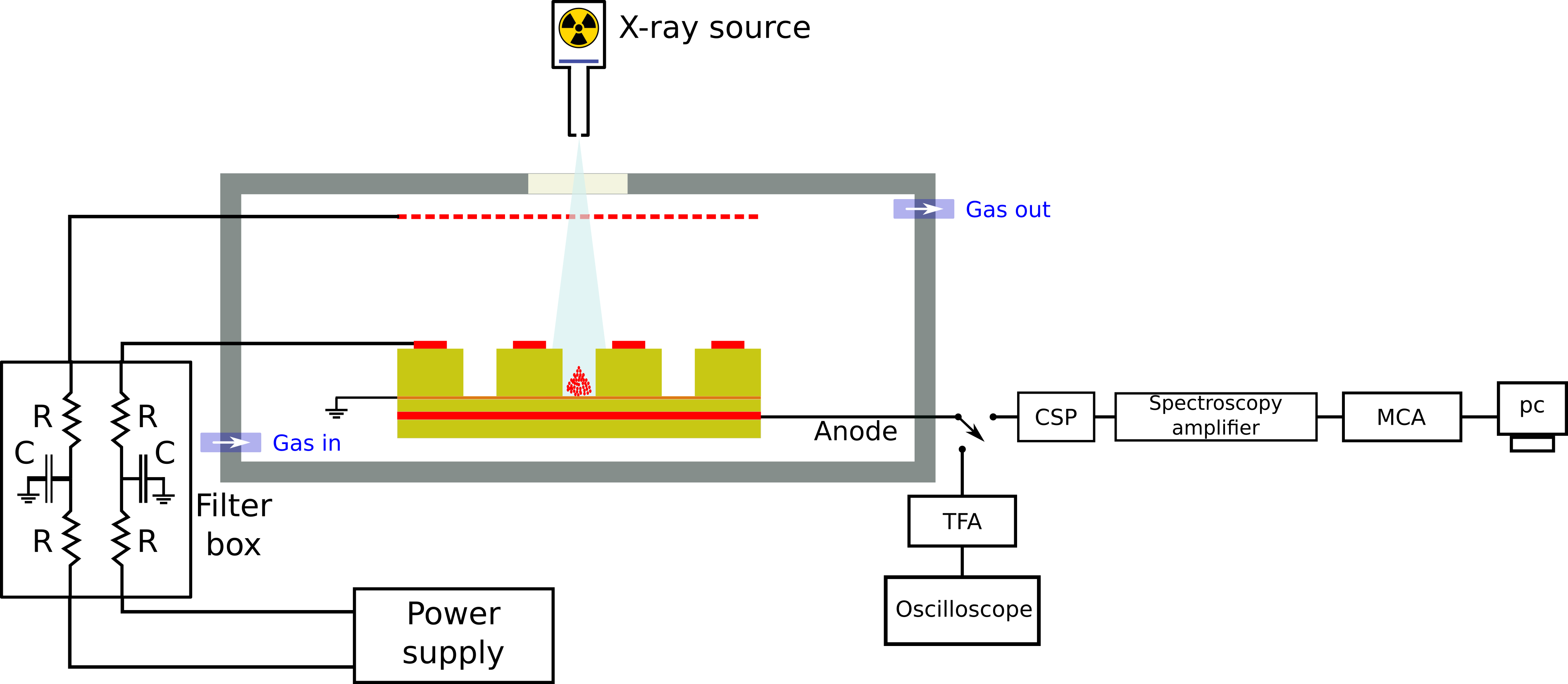}
	\caption{Schematic diagram of the detector assembly (not to scale). R $\sim$10 M$\Omega$, C $\sim$10 nF.
	\label{fig:expt_setup_schema}
	}
\end{figure}

A schematic diagram of the experimental setup is shown in Figure \ref{fig:expt_setup_schema} with an RWELL detector located within a small aluminum vessel.
Similar setup was used for the THWELL and RPWELL.
The chamber is constantly flushed with 20 cc/min Ne/5\%CH\textsubscript{4} at atmospheric pressure.
The detectors were irradiated with 8.04 keV X-rays from a Cu-target generator (Oxford instruments 5500). A few \textmu m thick Nickel filter suppresses low-energy Bremsstrahlung photons. A 20 cm long Al tube with a 0.5 mm diameter opening collimates the X-ray beam into the detector. The X-ray rate is set by the current of the X-ray generator tube, operated at 30 kV.

The WELL-top and the cathode are biased using a CAEN N1471 power supply via a low-pass filter and a 10 M$\Omega$ decoupling resistor so that the capacitance (10 nF) in the filter does not add to the discharge. The anode is maintained at ground potential. It is connected to either the electronic chain with a charge-sensitive pre-amplifier (CSP) (CANBERRA 2006), spectroscopic amplifier (ORTEC 471), Multi-Channel Analyzer (MCA) (Amptek 8000D), or the chain with a timing-filter amplifier (TFA) (ORTEC 474) and a digital oscilloscope (Keysight DSOX3034A). The second one was used to avoid saturation while reading signals from the anode.
The CSP signals were shaped using a spectroscopic amplifier. The case-specific connections will be described in section \ref{section:method}.
\subsection{Methodology}
\label{section:method}
\begin{figure}[htb]
 \centering
 \subfloat[a][]{
    \centering
    \includegraphics[width=0.485\textwidth, trim={0.0cm 0.0cm 1.5cm 1.5cm},clip]{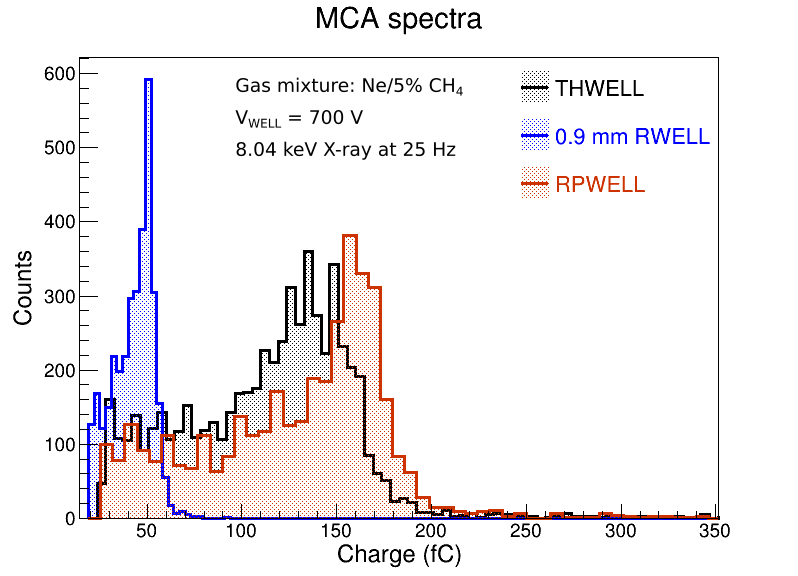}
    \label{fig:MCA_dist_3Det_700V}
 }
 \hfill
  \subfloat[b][]{
    \centering
    \includegraphics[width=0.485\textwidth, trim={0.0cm 0.0cm 2cm 1.5cm},clip]{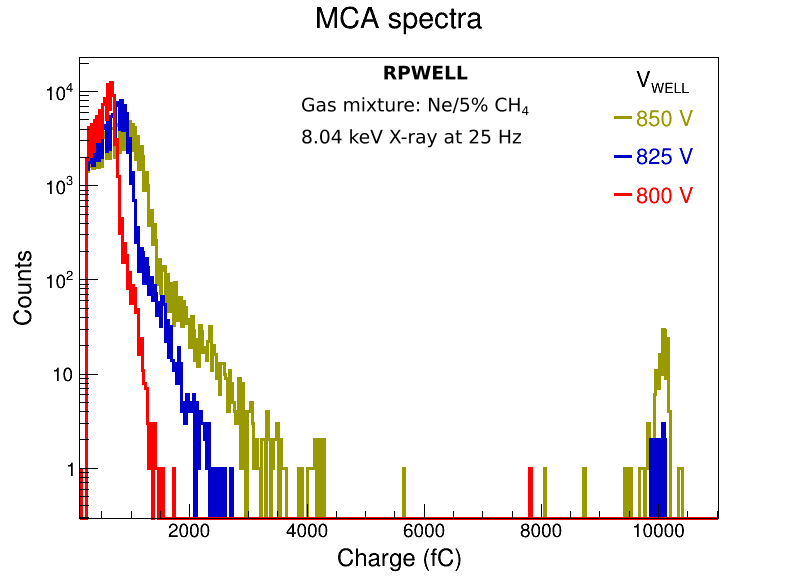}
    \label{fig:MCA_dist_RPWELL_log}
 }
 \caption{\protect\subref{fig:MCA_dist_3Det_700V} Calibrated MCA charge spectra from the three detectors at V\textsubscript{WELL} = 700 V, \protect\subref{fig:MCA_dist_RPWELL_log} MCA charge spectra of the RPWELL at higher voltages; the population at high charges, increasing with voltage, corresponds to saturation of the CSP by discharges.}
 \label{fig:MCA_dist_3Det_RPWELL}
\end{figure}

\begin{figure}[htb]
 \centering
 \includegraphics[width=\textwidth, trim={0.0cm 0.2cm 2.5cm 1.5cm},clip]{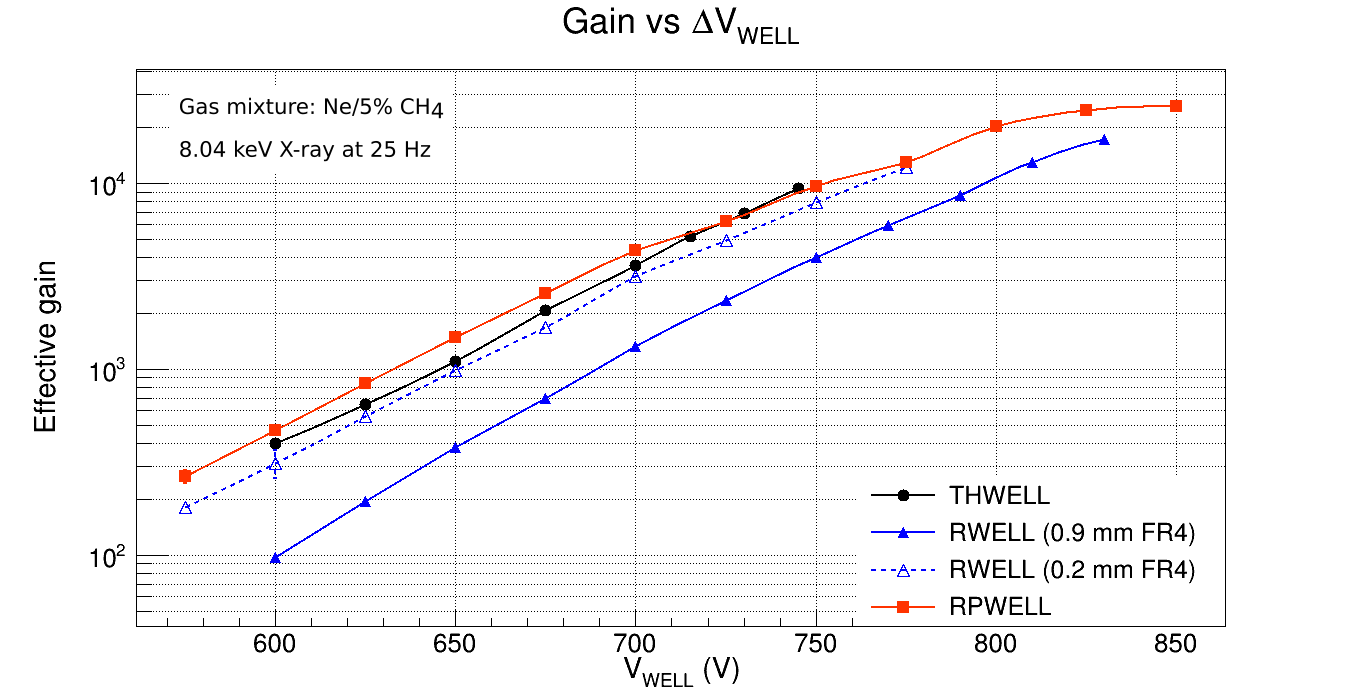}
 \caption{Variation of effective gain of the investigated detectors with the applied voltage at a source rate of 25 Hz.}
 \label{fig:gain_vs_dvWell_3Det}
 \end{figure}

Upon assembly of a detector setup and before any measurement, the system was flushed with the Ne/5\%CH\textsubscript{4} gas mixture for an hour at 20 cc/min, corresponding to three volume changes. Then the electrodes were biased, and the X-ray source was switched on to irradiate the detector. Similar to the observations in \cite{chargingUp_michael, gainStabilization_dan}, we see a fast gain-drop of the detectors followed by a slow increase due to charging up/down of the FR4 rims and walls. This effect stabilizes in about 4 hours from the start of irradiation and in about 15 minutes after any voltage change.
We waited for 12 hours after switching on the X-ray source to ensure gain stabilization. Likewise, we waited for 30 minutes after any change in detector voltage.
\paragraph{Gain characterization:}
The first step of the study involved the basic characterization of the detector in terms of charge spectrum and gain. The results were taken as a reference for the operating conditions in the discharge studies. To obtain the charge spectra, the anode was connected to the CSP/amplifier/MCA chain, which was calibrated by injecting a known charge at its input.
The spectra from the anode of the three detectors at V\textsubscript{WELL} = 700 V are shown in Figure \ref{fig:MCA_dist_3Det_700V}. 
The meaning of charge in these measurements is that of charge induced on the anode by the movement of avalanche charges inside the detector. This can be different from the total avalanche charge. For example, at the same voltage, a significantly lower charge was measured in the RWELL compared to the other two detector configurations.
%
The effective detector gain was calculated by dividing the induced charge extracted from the 8.04 keV peak of the spectrum by the primary charge deposited by the incident radiation (the term "effective" refers here to the induced nature of the signal).
Using HEED \cite{HEED}, the calculated average number of primary electrons produced by 8.04 keV photons in Ne/5\%CH\textsubscript{4} is 229, corresponding to a primary charge of $\sim$0.04 fC.
The effective gain of the three detectors as a function of V\textsubscript{WELL} is shown in Figure \ref{fig:gain_vs_dvWell_3Det}, for X-ray rate of $\sim$25 Hz.
Both the spectra and the effective gain curves reflect the fact that the induced charge of the THWELL and RPWELL detectors was similar, whereas a significantly lower induced charge was observed for the 0.9 mm thick RWELL. The latter can be understood from the Shockley-Ramo theorem \cite{shockley, ramo} and its extension for resistive electrodes \cite{riegler_signalResistive} which state that the magnitude of induced current on a readout electrode is proportional to its weighting field vector. Indeed, repeating the same measurement with 0.2 mm thick RWELL resulted in a gain similar to the one measured with the THWELL and RPWELL, as shown in Figure \ref{fig:gain_vs_dvWell_3Det}. 
The RWELL measurements reported in this work were conducted using the 0.9 mm thick FR4 as the substrate for the resistive layer.
Nevertheless, to reflect the actual avalanche charge, signals induced on the RWELL anode are calibrated to match the THWELL charge at the same voltage. The calibration parameters were obtained from an exponential fit to the THWELL gain curve.

The same logic of the weighting field does not explain the RPWELL gain curve. As seen in Figure \ref{fig:gain_vs_dvWell_3Det} and as reported in \cite{RPWELL}, the addition of a resistive plate has a small effect on the measured charge and corresponding effective gain, regardless of the plate thickness. To understand this unexpected behavior, further investigation is needed.
Similar to other gas-avalanche detectors, the effective gain of WELL structures increases exponentially with the applied voltage.
A maximum effective gain of about 9.5 $\times$ 10\textsuperscript{3} was reached in the THWELL at V\textsubscript{WELL} = 745 V, above which intense discharges at a high rate prohibited the detector operation. Due to the discharge quenching mechanism, the RWELL detector could operate at $\sim$2-fold higher gain and up to V\textsubscript{WELL} = 830 V. At higher voltages, the discharge probability was too high, and a proper charge spectrum could not be acquired.
For the RPWELL, no current fluctuations were observed, even at relatively high operation voltage values. Nevertheless, high-charge events saturating the CSP appeared, whose frequency was voltage-dependent. They are represented as an additional population around 10000 fC in the MCA spectrum in Figure \ref{fig:MCA_dist_RPWELL_log}. The RPWELL gain curve was found to deviate from exponential, saturating with voltage (similar to previous observations \cite{RPWELL}).
A maximum gain of about 3$\times$10\textsuperscript{4} was reached at 850 V, above which the acquired spectra were distorted. 

\begin{figure}[htb]
 \centering
 \subfloat[a][]{
    \centering
    \includegraphics[width=0.485\textwidth, trim={0.0cm 0.0cm 2cm 1.2cm},clip]{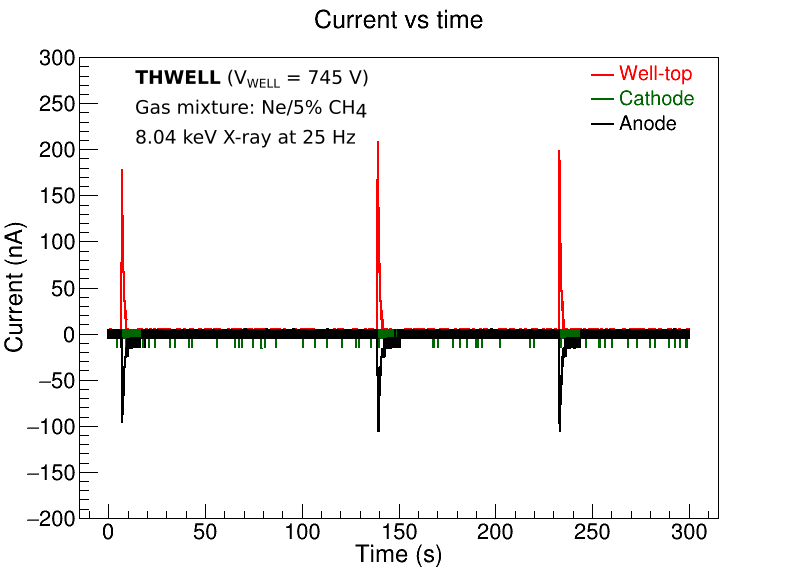}
    \label{fig:Imon_THWELL_disc}
 }
 \hfill
 \subfloat[b][]{
    \centering
    \includegraphics[width=0.485\textwidth, trim={0.0cm 0.0cm 2cm 1.2cm},clip]{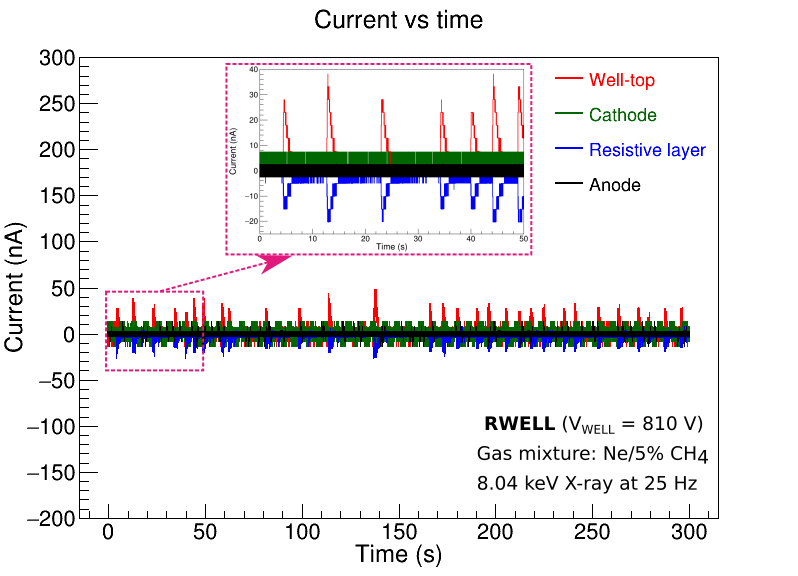}
    \label{fig:Imon_RWELL_disc}
 }
 \caption{The supplied currents to all the electrodes of \protect\subref{fig:Imon_THWELL_disc} a THWELL at V\textsubscript{WELL} = 745 V, \protect\subref{fig:Imon_RWELL_disc} a graphite RWELL at 810 V (zoomed view in the inset) with an effective gain around 10\textsuperscript{4}.}
 \label{fig:Imon_THWELL_RWELL}
\end{figure}

The effective RPWELL gain was found to have a strong dependence on the radiation rate \cite{RPWELL}; it is discussed in detail in appendix \ref{appendix:RPWELL_gain_vs_rate}.
To minimize the effects due to source rate, all following measurements were performed at an acquisition rate of $\sim$25 Hz.
\paragraph{Discharge characterization by power-supply currents:}
The standard method of discharge identification in a gas detector relies on the observation of the sudden rise of currents supplied to the electrodes. The power-supply current monitors of all electrodes were digitized using NI USB-6008 DAQ with Signal Express software at a sampling rate of 2 kHz. In normal operating conditions, the electrode currents had a zero value. The current measured along 5 minutes on all the THWELL detector electrodes is shown in Figure \ref{fig:Imon_THWELL_disc} at V\textsubscript{WELL} = 745 V, where discharges appeared occasionally.
The positive WELL-top current is accompanied by a simultaneous negative current on the anode, implying a current flow between the two during a discharge.
Pile-up of these spikes, appearing occasionally, indicates possible multiple-discharge events. The large current is accompanied by a significant (up to 20 V) voltage drop observed in the power supply.
The recovery time (2 seconds in the present case) is a property of the power supply itself.

The electrode currents measured in the RWELL in the presence of discharges are shown in Figure \ref{fig:Imon_RWELL_disc}. The occurrence of a discharge shows current spikes similar to the one observed with the THWELL, but the opposite polarity spikes are measured on the WELL-top and the resistive layer (instead of the anode in the THWELL). As also shown in \cite{Lior_WELLs}, the amplitudes of the WELL-top current spikes in RWELL are lower compared to the ones measured in a THWELL operated at the same effective gain (Figure \ref{fig:Imon_THWELL_disc}), demonstrating the effect of the resistive layer in quenching discharges.
The similar behavior between THWELL and RWELL points to the fact that the discharge mechanism is the same.
The lower amplitude of the negative polarity current spikes compared to the corresponding positive polarity spikes is an artifact of the power supply settings, which only affects the electrode biased at 0 V.
For both detectors, we could measure the discharge intensity as the total charge produced, i.e., the integral of the power supply's current spike at the WELL-top.

\begin{figure}[htb]
 \centering
 \subfloat[a][]{
    \centering
    \includegraphics[width=0.47\textwidth, trim={0.0cm 0.2cm 1cm 1.2cm},clip]{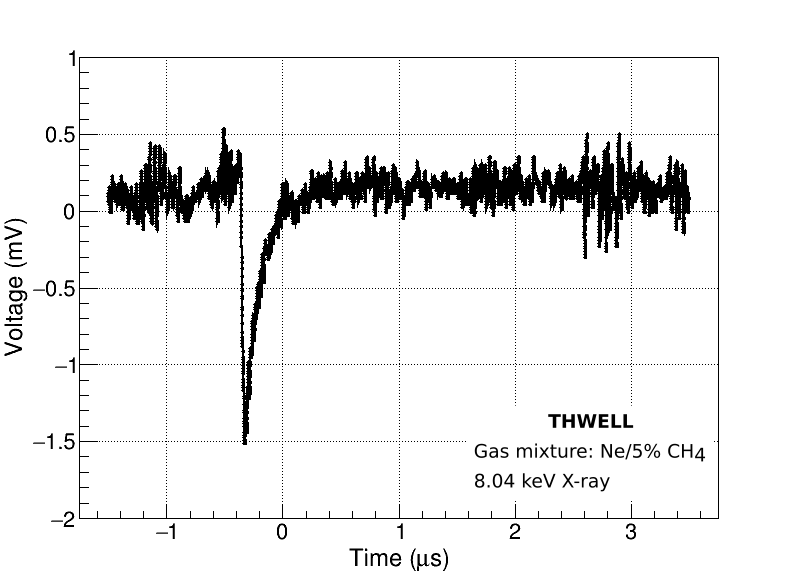}
    \label{fig:signalShapes_WELL_aval}
 }
 \hfill
 \subfloat[b][]{
    \centering
    \includegraphics[width=0.47\textwidth, trim={0.0cm 0.2cm 1cm 1.2cm},clip]{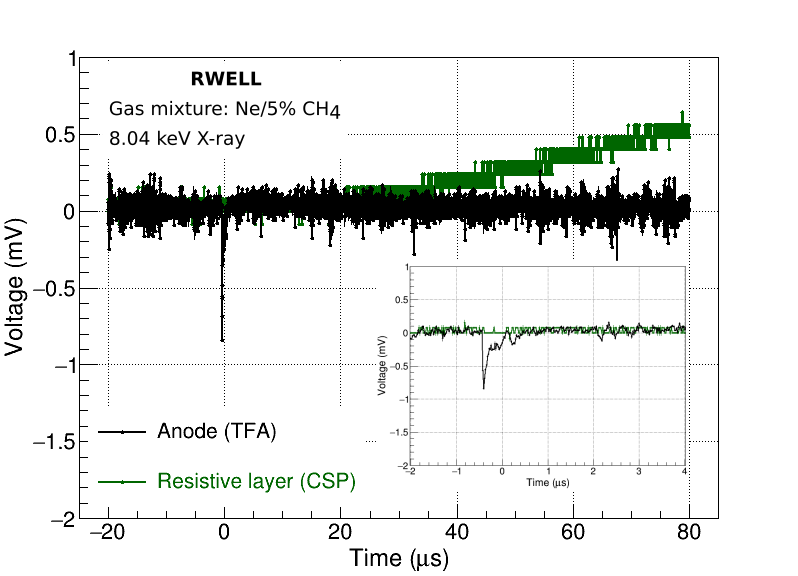}
    \label{fig:signalShapes_RWELL_aval_largeScale}
    }
 \\
 \subfloat[c][]{
    \centering
    \includegraphics[width=0.47\textwidth, trim={0.0cm 0.2cm 1cm 1.2cm},clip]{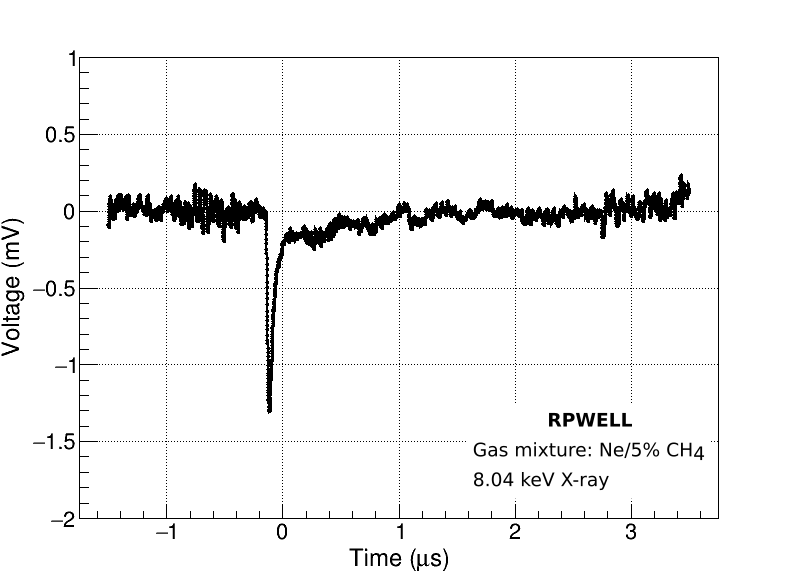}
    \label{fig:signalShapes_RPWELL_aval}
    }
 \caption{Typical avalanche-like signals from a \protect\subref{fig:signalShapes_WELL_aval} THWELL anode, \protect\subref{fig:signalShapes_RWELL_aval_largeScale} RWELL anode and resistive layer, \protect\subref{fig:signalShapes_RPWELL_aval} RPWELL anode. The anode signals are amplified and shaped by a TFA with 10 ns integration time and 150 \textmu s differentiation time. The signal from the resistive layer in RWELL is inverted and integrated by a CSP (refer to section \ref{section:setup}).}
 \label{fig:signalShapes_WELL_RWELL}
\end{figure}

The current fluctuations due to discharges in the RPWELL configurations were of the order of hundreds of pA to a few nA (compared to $\sim$ 200 nA and 40 nA in the THWELL and RWELL, respectively). While they were visible on the power-supply controller screen, they were below the sensitivity of the NI-DAQ recording system. Due to the limited sensitivity, the RPWELL detector was described as "discharge-free" in previous publications \cite{RPWELL, Luca_RPWELL_Ar, Shikma_RPWELL_in-beam, Shikma_THGEM_advances}.
\paragraph{Discharge characterization from induced signals:}
We used the electronic chain of the TFA and oscilloscope to monitor the anode pulses. 
A typical avalanche signal from the THWELL anode at normal operating conditions, amplified and shaped by the TFA, is shown in Figure \ref{fig:signalShapes_WELL_aval}.
The electrons (ions) produced in an avalanche drift towards the anode (WELL-top) and induce a negative (positive) signal of a few mV amplitude. The $\sim$50 ns rise time is attributed to the fast motion of electrons \cite{purba_THGEM_simulation}. The $\sim$\textmu s tail is due to the slower motion of ions away from the anode.

In the THWELL, RWELL, and RPWELL, the avalanche-induced anode signals have similar shapes \cite{RPWELL}. In the RWELL, an additional low-current was measured on the resistive layer. Its time profile was consistent with charge diffusion on the resistive layer. A typical case is shown in Figure \ref{fig:signalShapes_RWELL_aval_largeScale} where the anode signal is amplified by the TFA and the resistive layer one is integrated by the CSP. Considering the polarity inversion in the CSP, the signals from the anode and the resistive layer are of the same polarity.
The slow evacuation of the deposited electrons across the graphite layer to the ground has two main effects on the detector performance. During a gas breakdown, a large number of charges accumulate locally, causing a field reduction and consequently quenching the discharge energy. Also, the charge movement within the resistive material causes a voltage drop along its path to the ground. This is the cause of the rate dependence of the gain in resistive detectors \cite{RPC_highRate, resistiveDet_rateDependence}.

\begin{figure}[htb]
 \centering
 \subfloat[a][]{
    \centering
    \includegraphics[width=0.485\textwidth, trim={0.0cm 0.2cm 1cm 0.8cm},clip]{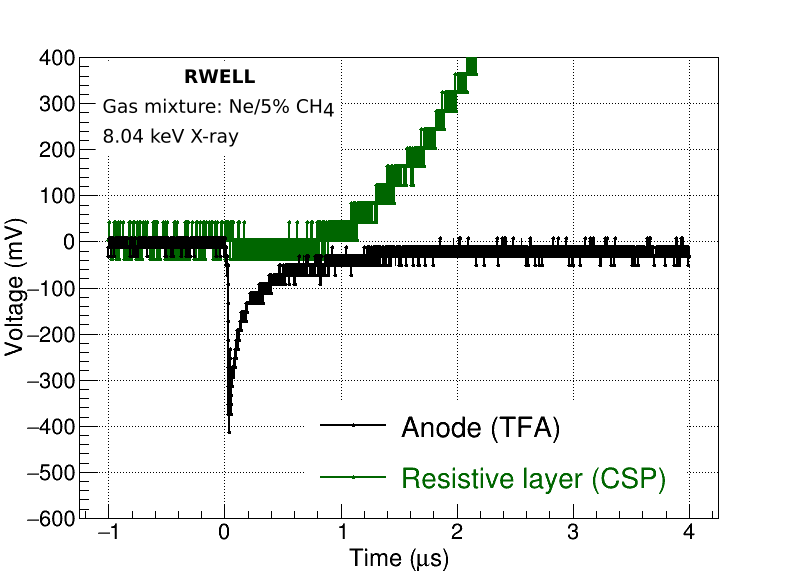}
    \label{fig:signalShapes_RWELL_disc_largeScale}
 }
 \hfill
 \subfloat[b][]{
    \centering
    \includegraphics[width=0.485\textwidth, trim={0.0cm 0.2cm 1cm 0.8cm},clip]{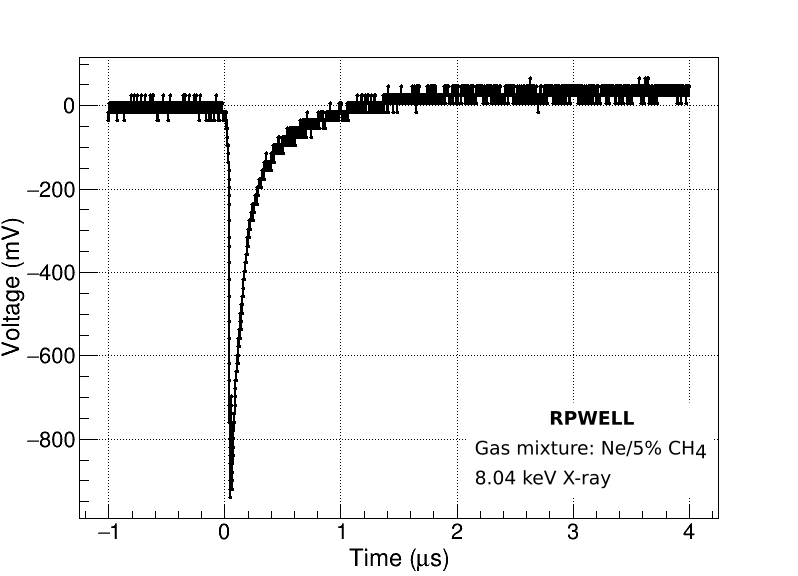}
    \label{fig:signalShapes_RPWELL_disc}
 }
 \caption{Typical discharge-like signals from \protect\subref{fig:signalShapes_RWELL_disc_largeScale} RWELL anode and resistive layer, \protect\subref{fig:signalShapes_RPWELL_disc} RPWELL. The anode signals are amplified and shaped by a TFA with 10 ns integration time and 150 \textmu s differentiation time. The signal from the resistive layer in RWELL is inverted and integrated by a CSP.}
 \label{fig:signalShape_chargeDistribution_RPWELL}
\end{figure}

Around their maximal operation voltages, high-amplitude signals of the order of a few hundreds of mV are induced on the RWELL and RPWELL anodes. These pulses have similar rise times as that of the avalanche-induced pulses. In the RWELL, these pulses are correlated with discharges identified by the power supply's current spikes. A typical high-charge pulse in the RWELL is shown in Figure \ref{fig:signalShapes_RWELL_disc_largeScale}. %
In the case of RPWELL, the appearance of these large-amplitude pulses is correlated with a few nA fluctuations in the WELL-top power-supply current. A typical induced pulse for a discharge-like event in RPWELL is shown in Figure \ref{fig:signalShapes_RPWELL_disc}.
Discharges occurring in the THWELL saturate the TFA (input range $\pm$1 V), and no meaningful signal could be acquired.

The high-amplitude pulses were used to trigger discharge events in the two resistive configurations. The amplitude of these discharge-like pulses after TFA was measured using the oscilloscope (with 50 $\Omega$ input impedance). It was calibrated to charge following the method described in \cite{luca_detectorEmulator}.

For the three detector configurations, the discharge probability was estimated as the number of discharges acquired over the acquisition time and the source rate.
\section{Results}
\label{section:results}
\begin{figure}[htb]
 \centering
 \subfloat[a][]{
    \centering
    \includegraphics[width=0.48\textwidth, trim={0.0cm 0.2cm 1cm 1.5cm},clip]{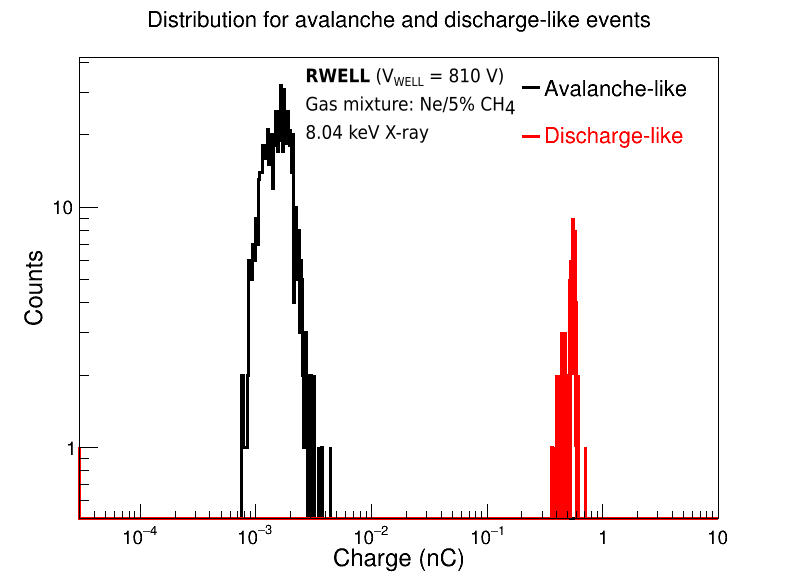}
    \label{fig:signalShapes_chargeDistribution_RWELL_aval_disc}
 }
 \hfill
 \subfloat[b][]{
    \centering
    \includegraphics[width=0.48\textwidth, trim={0.0cm 0.0cm 1cm 1.5cm},clip]{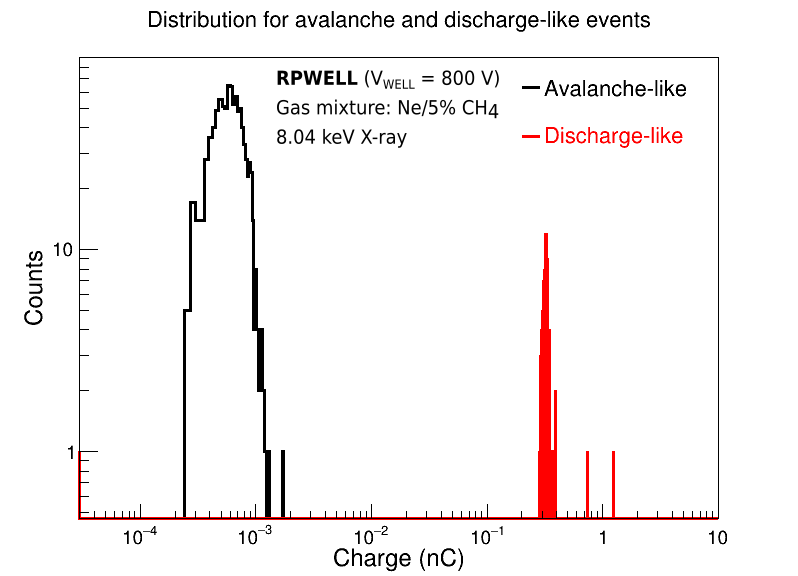}
    \label{fig:signalShapes_chargeDistribution_RPWELL_aval_disc}
 }
 \caption{Charge distribution in avalanche and discharge-like events calculated from \protect\subref{fig:signalShapes_chargeDistribution_RWELL_aval_disc} RWELL pulses at V\textsubscript{WELL} = 810 V, \protect\subref{fig:signalShapes_chargeDistribution_RPWELL_aval_disc} RPWELL pulses at V\textsubscript{WELL} = 800 V (pulse acquisition using TFA with 20 ns integration time and 150 \textmu s differentiation time).}
 \label{fig:signalShape_chargeDist_aval_disch}
\end{figure}

Following the methodology discussed in detail in section \ref{section:method}, the pulses induced on the RWELL and RPWELL anodes were measured with the TFA. The total induced charge was estimated from the pulse heights, taking into account the RWELL calibration. The charge distributions for the RWELL and RPWELL are depicted in Figures \ref{fig:signalShapes_chargeDistribution_RWELL_aval_disc} and \ref{fig:signalShapes_chargeDistribution_RPWELL_aval_disc}, respectively. 
In both cases, two populations separated by about three orders of magnitude are observed. 
The first ones (black) correspond to the regular, avalanche-induced pulses. They are observed in all three detector configurations in their proportional operation mode. 
Avalanche-like pulses with a total charge below a few hundred fC are not well-separated from the noise and could not be read out.
The second populations (red curve) correspond to the charge induced on the readout anode by discharge events. Its typical magnitude is of the order of hundreds of pC.

The end-point of the avalanche-like distributions ($\sim$a few pC) corresponds to a few 10\textsuperscript{6} electrons, in agreement with the critical charge reported for THGEM and THGEM-based WELL structures \cite{Shikma_instabilities_THGEM, peskov_THGEM_RICH, Marco_THGEM_NeCH4}.
This strengthens the claim that the large-amplitude pulses are induced by discharges. In the following, discharge-like events are selected by setting a 70 pC threshold.

\begin{figure}[htb]
 \centering
 \subfloat[a][]{
    \centering
    \includegraphics[width=0.48\textwidth, trim={0.0cm 0.2cm 1cm 1.2cm},clip]{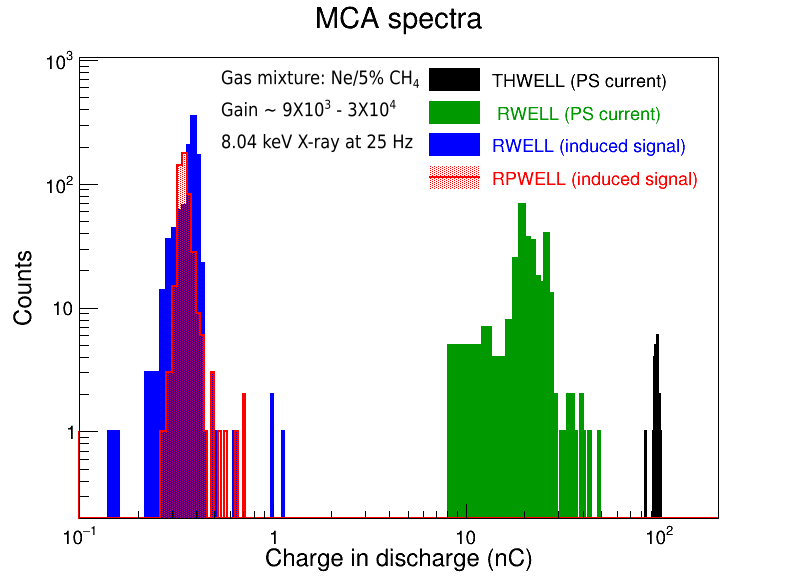}
    \label{fig:hist_Qdischarge_3Det}
 }
 \hfill
 \subfloat[b][]{
    \centering
    \includegraphics[width=0.48\textwidth, trim={0.0cm 0.0cm 1cm 0.cm},clip]{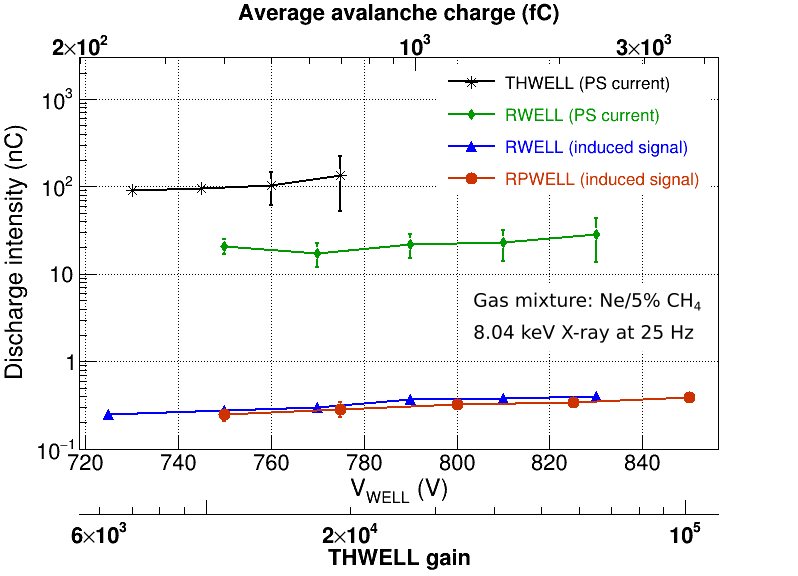}
    \label{fig:Qdischarge_vs_gain_3Det}}
\caption{\protect\subref{fig:hist_Qdischarge_3Det} Distribution of charge in discharge for the three investigated detectors. The bias voltage and the effective gain values are: THWELL (745 V, G = $9.4\times10^{3}$), RWELL (810 V, G = $1.3\times10^{4}$), RPWELL (825 V, G = $2.5\times10^{4}$),
\protect\subref{fig:Qdischarge_vs_gain_3Det} variation of discharge intensity with the applied voltage of the three investigated detectors at a source rate of 25 Hz. The extrapolated induced charge and the effective gain of THWELL at the corresponding voltages are also shown. The 70-fold difference in the RWELL discharge intensities measured from the two methods are discussed below.}
 \label{fig:dischargeIntensity_3Det}
\end{figure}

The distribution of the charge induced on the readout anode in discharge-like pulses is shown in Figure \ref{fig:hist_Qdischarge_3Det} for the RWELL (blue) and the RPWELL (red). The charge estimated from the peaks of the current supplied to the WELL-top electrode in the THWELL (black), and RWELL (green) are also shown.
To ensure sufficient statistics, the measurements were taken at an effective gain in the range $9\times10^{3} - 3\times10^{4}$. The mean of the Gaussian fits of these distributions and their respective errors are shown in Figure \ref{fig:Qdischarge_vs_gain_3Det} as a function of three different parameters: (1) the voltage applied to the detectors, (2) the THWELL gain, and (3) the average avalanche charge. 
Several observations can be made. For all the investigated configurations, the discharge intensity (measured charge) has a very small dependency on the gain.
The energy released in the discharge of a THWELL is an order of magnitude larger than that released in the discharge of an RWELL. For the latter, the charge measured on the anode (induced pulses) is about two orders of magnitude smaller than that measured on the WELL-top (power supply current). This large difference should be further investigated. It could suggest that the two quantities are governed by two different physics processes. The charge estimated from the power-supply current corresponds to the discharge of the WELL-top/resistive-layer capacitor. The current, in this case, flows through a conductive path between the WELL-top and the resistive layer and, thus, does not induce a signal on the anode.
The charge measured on the anode could be induced by the large number of charges moving within the gaseous medium after crossing the critical charge limit and before the streamer connects the WELL-top and the resistive layer. As expected, this amount should not change much for the RWELL and RPWELL configurations. In both cases, the discharge intensity is quenched. For the THWELL, this is not the case, and the signal caused by the large charge released in the discharge hides any other phenomenon.
\begin{figure}[htb]
 \centering
 \includegraphics[width=\textwidth, trim={0.0cm 0.0cm 1cm 0.0cm},clip]{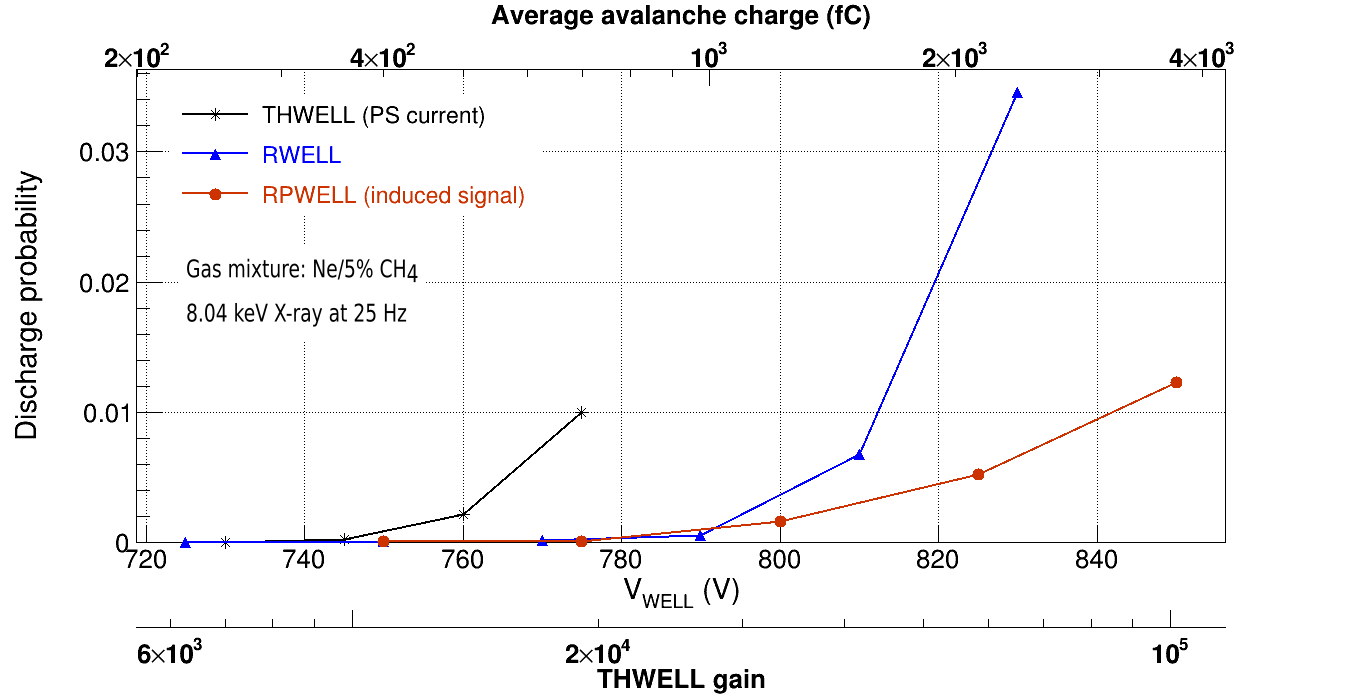}
\caption{Variation of discharge probability with the applied voltage of the three investigated detectors at a source rate of 25 Hz. The extrapolated induced charge and the effective gain of THWELL at the corresponding voltages are also shown.}
\label{fig:dischargeRate_vs_gain_3Det}
\end{figure}

The discharge probability of the three detectors is shown in Figure \ref{fig:dischargeRate_vs_gain_3Det} as a function of the voltage, the THWELL gain, and the average avalanche charge.
The discharge probability of the THWELL rises earlier than the two resistive configurations. The rise in the discharge probability of the RPWELL is less steep than the other two.
Given the low event rate, the possibility of gain drop due to a single discharge being the reason for the suppression of consequent discharges in the two resistive configurations is excluded; this suggests that another mechanism could mitigate the occurrence of discharges in these configurations, which is slightly more pronounced in the RPWELL.
The resistive detectors could be operated with discharge probability up to about 0.02, above which the measured spectrum became distorted. This corresponds to maximal gain values of about a few $10^{4}$. Due to the high intensity of the THWELL discharges, the operation of this configuration was limited to a discharge probability of about 0.002, corresponding to a gain of 10\textsuperscript{4}.
\section{Summary}
\label{section:summary}
In this work, we have shown that discharges occur in WELL-based detectors with and without resistive anodes. Nonetheless, they take different forms in different configurations, and consequently, they have different effects on the detector performance. 

In the case of the non-resistive THWELL, once the critical charge limit is crossed and the gas breaks down, a direct conductive path is formed between the WELL-top and the anode (ground), resulting in an intense discharge of the energy stored between the two. These violent discharges limit the voltage that can be applied to the detector, its gain, and eventual dynamic range.

In the RWELL and RPWELL configurations, there is no direct path between the WELL-top and the ground. Instead, the electrons are forced to flow through a resistive path. The resistances seen by the respective evacuated charges are about 16 M$\Omega$ and hundreds of G$\Omega$ for the RWELL and RPWELL investigated in this work (see appendix \ref{appendix:RPWELL_gain_vs_rate}), respectively. The slow charge evacuation via the resistive path decreases the local electric field, thus quenching the energy released in a discharge.

For all investigated configurations, no current was measured on the cathode during a discharge. In the THWELL configuration, current spikes of opposite polarity were measured on the WELL-top and the anode. In the RWELL configuration, the opposite polarity current spikes were measured on the WELL-top and the resistive layer. These were accompanied by large pulses on the anode - two orders of magnitude larger than avalanche-like pulses. Similar to the RWELL, the occurrence of discharges in the RPWELL induces large pulses on its anode. This is accompanied by small current fluctuations on the WELL-top, below the 5 nA sensitivity of the recording instrument. Similar to the observations made in \cite{Lior_WELLs}, the discharge intensity in the RWELL measured from the current supplied to the WELL-top electrode was found to be five times smaller than that in THWELL. Moreover, the intensity of the discharge measured on the anode was two orders of magnitude lower.

The standard discharge identification method by monitoring electrode currents (e.g., in \cite{Lior_WELLs, Lior_DHCAL, Shikma_instabilities_THGEM}) is useful when the streamer developed after the gas breakdown connects two electrodes and discharges the effective capacitor defined by them.
In detectors with resistive anodes, where the readout electrode is shielded from the gaseous medium (e.g., in the case of RPWELL, RPC \cite{RPC_streamer} and other resistive gaseous detectors), gas breakdown induces large pulses on the readout element, but no current may be detected by standard current monitoring devices. The newly presented method will be useful in identifying gas breakdown in such cases.
\appendix
\section{The effective resistance of the glass plate}
\label{appendix:RPWELL_gain_vs_rate}
The effective resistance of the glass plate was estimated from the gain variation at different source rates. The total current due to the avalanche is given by $i = r \times Q_{0} \times G$, where $r$ is the event rate, $Q_{0}$ is the average primary charge, and $G$ is the effective gain of the detector.  The flow of this current through the bulk of the resistive plate creates a voltage drop such that the effective voltage difference between the WELL-top and the resistive layer is given by $V_{eff}$ = $V_{WELL}$ - $R \times r \times Q_{0} \times G$, where $R$ is the effective resistance. Smaller $V_{eff}$ is obtained for a higher source rate resulting in lower detector gain. 
\begin{figure}[htb]
 \centering
 \subfloat[a][]{
    \centering
    \includegraphics[width=0.485\textwidth, trim={0.0cm 0.0cm 1cm 1.5cm},clip]{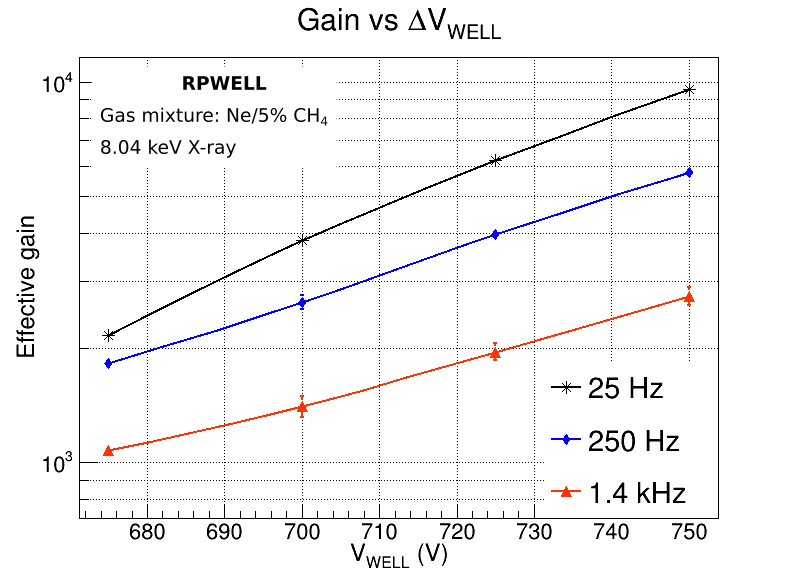}
    \label{fig:gain_vs_dvWell_RPWELL_differentRates}
 }
 \hfill
 \subfloat[b][]{
    \centering
    \includegraphics[width=0.485\textwidth, trim={0.0cm 0.0cm 1cm 1.6cm},clip]{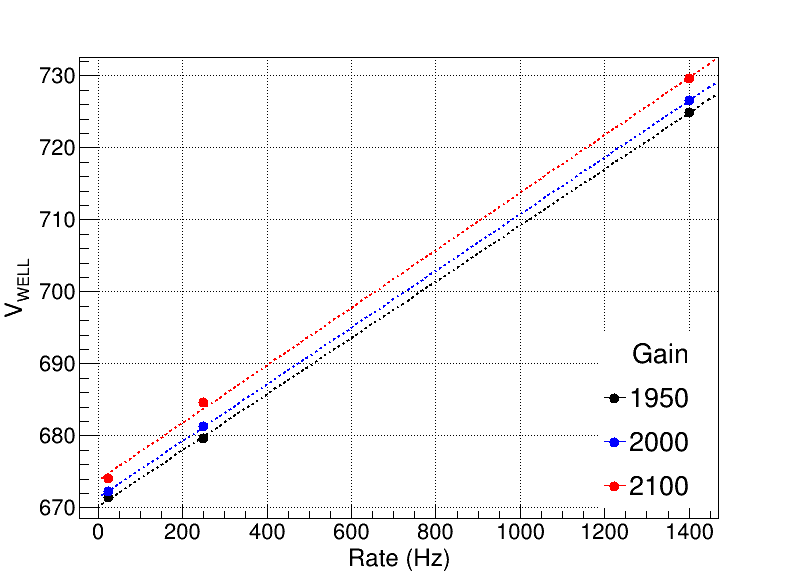}
    \label{fig:dvWell_vs_sourcerate_RPWELL_differentGains}
 }
 \caption{\protect\subref{fig:gain_vs_dvWell_RPWELL_differentRates} Variation of the effective gain of the RPWELL with the applied voltage at different X-ray acquisition rates, \protect\subref{fig:dvWell_vs_sourcerate_RPWELL_differentGains} plot of required voltage vs X-ray rate to reach a given effective gain of the RPWELL.}
 \label{fig:gain_RPWELL_rateDependance_dvWell_forFixedGain}
\end{figure}

The effective gain of the RPWELL as a function of voltage is shown in Figure \ref{fig:gain_vs_dvWell_RPWELL_differentRates} for different X-ray rates. The corresponding V\textsubscript{WELL} needed to ensure fixed effective gain at different rates is depicted in Figure \ref{fig:dvWell_vs_sourcerate_RPWELL_differentGains}. These curves are fitted to straight lines in which the slope is related to the resistance as follows:
$$
\text{R = } \frac{\Delta V_{WELL}/\Delta r}{Q_{0} \times G}
$$
The resistance was found to be of the order of 500 G$\Omega$; it is in reasonable agreement with the direct calculation of the effective resistance under the participation of a single hole (700 G$\Omega$) or two holes (350 G$\Omega$) in the charge evacuation process.
\acknowledgments
This work was supported by Grant No. 3177/19 from the Israeli Science Foundation (ISF), The Pazy Foundation, and by the Sir Charles Clore Prize.
We are grateful to Prof. Amos Breskin and Dr. David Vartsky for reviewing the manuscript and their helpful suggestions during the work. Thanks to our colleagues, Andrea Tesi, Darina Zavazieva and Dan Shaked Renous for their cooperation and many helpful discussions.
\end{document}